\documentclass[twocolumn,showpacs,preprintnumbers,amsmath,amssymb]{revtex4}

\usepackage{graphicx}
\usepackage{dcolumn}
\usepackage{bm}

\begin{document}

\title{Electronic structure and magnetism of Mn doped GaN}

\author{B. Sanyal, O. Bengone and S. Mirbt} 
\affiliation{Department of Physics, Uppsala University, Uppsala, Sweden}

\date{\today}

\begin{abstract}
Mn doped semiconductors are extremely interesting systems 
due to their novel magnetic properties suitable for the spintronics applications. It has been
shown recently by both theory and experiment 
that Mn doped GaN systems have a very high Curie temperature compared to that
of Mn doped GaAs systems. To understand the electronic and magnetic properties, we have studied
Mn doped GaN system in detail by a first principles plane wave method.
We show here the effect of varying Mn concentration on the electronic and magnetic properties.
For dilute Mn concentration, $d$ states of Mn form an impurity band completely separated
from the valence band states of the host GaN. This is in contrast to the Mn doped GaAs system where Mn
$d$ states in the gap lie very close to the valence band edge and hybridizes strongly with 
the delocalized valence band states.
 To study the effects of electron correlation, LSDA+U calculations have been performed. 
 Calculated exchange interaction in (Mn,Ga)N is short ranged in contrary to that in (Mn,Ga)As 
where the strength of the ferromagnetic coupling between Mn spins is not decreased substantially for
 large Mn-Mn separation.
Also, the exchange interactions are anisotropic in different crystallographic directions due to the 
presence or absence of connectivity between Mn atoms through As bonds.  

\end{abstract}

\vspace{20mm}

\maketitle

\section{Introduction}
Diluted magnetic semiconductors (DMS) are considered to be potential candidates for 
 present and future technological applications in semiconductor spintronics \cite{bookdms}.
 During the last decade, there have been numerous experimental and theoretical
studies of II-VI, III-V and IV-VI DMS. Among the III-V DMS, Mn doped GaAs system has been studied
 rigorously for the last few years. 
 This system shows a Curie temperature ($T_{C}$) of 110 K for
a Mn doping concentration of 10 $\%$. Despite several attempts, $T_{C}$ couldn't be raised 
beyond 175 K. Recently, there have been reports of some room temperature
DMS. They include Mn doped GaP \cite{mngap}, Mn doped chalcopyrite CdGeP$_{2}$
\cite{chalco}, Mn doped GaN \cite{sasaki,reed} etc. The origin of ferromagnetism in these compounds
is still under debate \cite{origin}. 

Dietl {\it et al.}\cite{dietl} predicted theoretically a high Curie temperature ($\sim$ 400 K) 
for Mn doped GaN (5 $\%$ Mn).
Their theory was based on a mean field model of hole mediated ferromagnetism.
As the Curie temperature for a Mn doped GaAs system is comparatively lower, the theoretical prediction
for a higher T$_{c}$ drew much attention. Sasaki {\it et al.} \cite{sasaki} 
grew wurtzite Mn doped
GaN films by the molecular beam epitaxy method. Magnetic measurements showed a very high Curie
temperature of about 940 K. They ruled out the possibility of phase segregation of some
ferromagnetic compound e.g. MnGa and Mn$_{4}$N which also have high Curie temperatures. 
Room temperature ferromagnetism in Mn doped GaN was also observed by Reed {\it et al.} \cite{reed}.
Deep level optical spectroscopy measurements \cite{korotkov} show that Mn forms a deep acceptor
level at 1.42 eV above the valence band maximum for GaN doped with small concentration of Mn.
It is to be noted that Mn forms an acceptor level at 0.11 eV above the valence band maximum in
 the case of Mn doped GaAs. So, 
the overlap of the Mn d-states with the valence band is rather strong in Mn doped GaAs compared to Mn
 doped GaN.

Recently, there have been a few first principles electronic structure calculations of Mn doped
GaN systems.  Fong {\it et al.}\cite{fong} performed electronic structure calculations of Fe and Mn doped 
GaN using the tight-binding linearized muffin tin orbital (TB-LMTO) method. 
Sato and Katayama-Yoshida \cite{sato} performed KKR-CPA (Korringa-Kohn-Rostoker-Coherent Potential Approximation)
calculations to study the relative stabilities of ferromagnetic and spin glass phases. They showed
that for a low concentration of Mn, ferromagnetism is favored whereas for the high concentration,
the spin glass phase is stable. 
 The disordered local moment model was assumed to describe the spin glass phase. They
explained the origin of ferromagnetism in these systems by a competition between double exchange
and superexchange interactions. Kulatov {\it et al.} \cite{kulatov} 
studied electronic, magnetic and optical properties of 
zinc-blende (Mn,Ga)N for different concentrations
of Mn by the TB-LMTO method in a supercell approach.
 Anomalous exchange interactions in III-V DMS were 
 found from calculations by Schilfgaarde and Mryasov \cite{schilf}. They predicted aggregation
of magnetic nanoclusters inside the III-V host.
Kronik {\it et al.} \cite{kronik} considered (Mn,Ga)N in wurtzite structure and performed 
electronic structure calculations using a plane wave pseudopotential method.  
In a recent preprint \cite{alessio}, the different origins of ferromagnetism in (Mn,Ga)As and
(Mn,Ga)N systems were discussed. The authors pointed out from self-interaction corrected (SIC) 
pseudopotential calculations that (Mn,Ga)N is characterized by localized
Mn 3$d$ states with a strong self-interaction. In (Mn,Ga)As, $d$ states are weakly correlated and are 
rather delocalized being strongly hybridized with As $p$ states. 
In this communication, we attempt to understand the electronic structure and magnetic interactions
in Mn doped GaN and GaAs systems.
The motivation of this paper is twofold : (a) to investigate electronic structure 
and magnetism of Mn doped GaN system in detail and (b) to have a comparison
with Mn doped GaAs system.
The paper is organized as follows : In the next section, we describe the computational details. In the
results section, a subsection describes the electronic structure and magnetism of the (Mn,Ga)N with varying
 Mn concentration in the local spin density approximation (LSDA). Then we present results from LSDA+U calculations.
 Finally, we show a comparison of exchange interactions in (Mn,Ga)N and (Mn,Ga)As systems. 
\section{Computational details}
GaN can be grown both in the zinc-blende and wurtzite structures. But, usually, Mn is doped in a
wurtzite GaN host \cite{sasaki}.
In our calculations, we have considered the wurtzite structure. 
Also, for a comparison, we show calculations for the zinc-blende structure.
Experimental lattice parameters such as 
a=3.189 \AA \ and c=5.185 \AA \ with a c/a ratio of 1.626 were 
taken for the calculations in the wurtzite structure.  
Results from the atomic relaxations revealed that the nearest neighbor bond-lengths between Mn
 and N change only  
by 3 $\%$ compared to that of the bulk GaN. This is in agreement with the results of 
 Kronik {\it et al.} \cite{kronik}.
 In general, Mn doping in substitutional site results in a small relaxation \cite{msm} 
of the nearest neighbor anions around it.

Calculations have been performed by an {\it ab-initio} plane wave
code (VASP) \cite{vasp}. Vanderbilt \cite{vander} 
type ultrasoft pseudopotentials were used for the LSDA calculations.  
LSDA+U calculations were done in the projector augmented wave (PAW) \cite{paw} method as implemented by Bengone
{\it et al.} \cite{bengone}.
 Ga 3d orbitals were included in the basis set of the Ga pseudopotential. 
 A kinetic energy cut-off of 350 eV was
used for the plane waves included in the basis set. Ceperley and Alder \cite{ca} exchange-correlation  
functional parameterized by Perdew and Zunger \cite{pz} was considered within LSDA.
 We have also checked that the results obtained 
within GGA (generalized gradient approximation) \cite{gga} are similar.
 A 8x8x6 k-points grid was used in the Monkhorst Pack scheme \cite{mp} for small supercells. 
For the largest supercell considered, a 2x2x1 grid was used.
 Local properties such as local density of states and local magnetic moments
were calculated by projecting the wave functions onto spherical 
harmonics \cite{eichler}. The radii chosen for the projection were
 1.31, 1.21 and 0.74 \AA \ for Mn, Ga and N respectively.

We have modeled the system using different supercell sizes to simulate different Mn concentrations.
 For wurtzite structure, 
supercells having 4, 8, 16, 32, 72 and 108 atoms were used to model a composition Mn$_{x}$Ga$_{1-x}$N 
for $x$=0.5, 0.25, 0.125, 0.0625, 0.028 and 0.018 respectively. 
For zinc-blende structure, we used a 64 atom-cell to simulate a Mn concentration of $x$=0.03125. 
 It has been found experimentally that Mn occupies the Ga-site \cite{expt_jap}. 
Therefore, in our calculational unit cell, 1 Ga atom was substituted by a Mn atom.

\section{Results}
\subsection{Calculations within LSDA}
In a III-V semiconductor, a cation vacancy creates 3 holes in the valence band leaving anion
dangling bonds. When Mn occupies the cation site, it donates 3 electrons to fulfill the bonding.
Mn is left with 4 unpaired $d$ electrons which give rise to 4 $\mu_{B}$/Mn atom. In a realistic 
 situation, there
can be compensating donors e.g. As antisites and interstitial Mn atoms \cite{larsprb}, 
present in the system to increase or decrease the magnetic moments. But as we are 
dealing with ideal uncompensated systems, we always obtain a magnetic moment of 
4 $\mu_{B}$/Mn atom for substitutional Mn. 

\begin{figure}
\includegraphics[width=0.45\textwidth,height=0.38\textheight,angle=270]{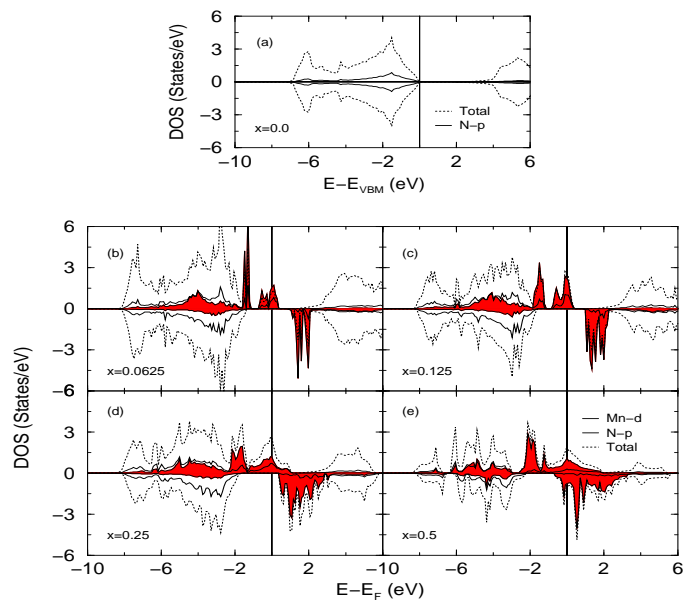}
\caption{Spin resolved density of states of (Mn$_{x}$Ga$_{1-x}$)N in the wurtzite structure. 
Data are shown for 
(a) pure GaN (b) 6.25 $\%$ Mn in GaN (c) 12.5 $\%$ Mn
(d) 25 $\%$ Mn (e) 50 $\%$ Mn. In (a), total DOS/cell as well as the $p$-DOS of N have been shown
whereas in (b) to (e), $d$-DOS of Mn (in shade), $p$-DOS of N and total DOS have been plotted. Energies are
plotted with reference to valence band maximum (VBM) of GaN in (a) and Fermi energies ($E_{F}$) in (b) to (e).}
\end{figure}

In Fig. 1(a-e), we show the density of states (DOS) for various concentrations($x$) of Mn in 
Mn$_{x}$Ga$_{1-x}$N in the wurtzite structure. In (a), the DOS of the undoped 
GaN is presented. The large band gap is evident from the figure. The calculated direct band gap at the 
$\Gamma$ point is 1.9 eV which is underestimated compared to the experimental band gap of $\sim$ 3.4 eV. This
 well known underestimation is inherent in the formulation of density functional theory 
and is well documented in existing literature.
 From Fig. 1(b) to 1(e), we show the DOSs with increasing $x$.
 The Mn impurity $d$ peak in the band gap 
is away from the valence band for small $x$. As the concentration
increases, this peak is broadened due to the overlap of Mn $d$ wavefunctions 
and the gap between the impurity peak
and valence band edge vanishes. The total DOS/cell gradually regains the shape of the DOS of undoped
system as $x$ is reduced. 
However, up to $x$=0.25, the Fermi level cuts only the
spin up DOSs. As there is no state at Fermi level for the spin down channel, we obtain a half metallic solution
giving rise to an integer magnetic moment of 4 $\mu_{B}$/Mn atom. For $x$=0.5 shown in Fig. 1(e), 
 the system behaves like a ferromagnetic metal with a reduced magnetic moment. One noticeable difference
between Mn doped GaAs and Mn doped GaN systems is the position of Mn impurity band in the gap. 
In (Mn,Ga)As, Mn impurity band almost
merges with the top of the valence band (approximately 0.1 eV above the valence band edge)
whereas in (Mn,Ga)N, the Mn impurity band is separated from the valence band by 0.56 eV. 
The width of this impurity band is 0.94 eV.
As the spin polarized Mn impurity band is distinctly in the gap, 
the valence band is less spin polarized in (Mn,Ga)N compared to (Mn,Ga)As.
Spin polarization ($\epsilon_{VBM/CBM}^{\uparrow}-\epsilon_{VBM/CBM}^{\downarrow}$)
of the valence and conduction bands at the $\Gamma$ point is 0.05 eV and -0.42 eV respectively 
for (Mn$_{0.0625}$Ga$_{0.9375}$)N. $\epsilon$ is the eigenvalue and VBM and CBM represent valence band 
maximum and conduction band minimum respectively.
\begin{center}
\begin{table}
\caption{ Projected charges and
magnetic moments (in $\mu_{B}$). Mn-d$^{Q}$ indicates charge in the
the sphere around Mn for d electrons. Mn$_{mom}$ and N$_{mom}$
indicate projected magnetic moments inside Mn and N spheres respectively.}
\begin{tabular}{|l|l|l|l|}
\hline \hline 
$x$ in Mn$_{x}$Ga$_{1-x}$N         & Mn-d$^{Q}$ &  Mn$_{mom}$ & $\frac{1}{4}\sum_{nn}$ N$_{mom}$ \\
\hline
0.018                    & 5.04        & 3.40     &  0.016 \\         
\hline
0.028			 & 5.04        & 3.41      & 0.016      \\ 
\hline
0.0625			 & 5.05        & 3.43      & 0.01      \\
\hline
0.125			 & 5.05        & 3.42      & 0.015      \\
\hline
0.25			 & 5.04        & 3.47      & 0.02       \\
\hline
0.50			 & 5.09        & 2.61      & -0.02      \\ 
\hline \hline
\end{tabular}
\end{table}
\end{center}
\begin{figure}[!htp]
\includegraphics[width=0.45\textwidth,height=0.3\textheight,angle=270]{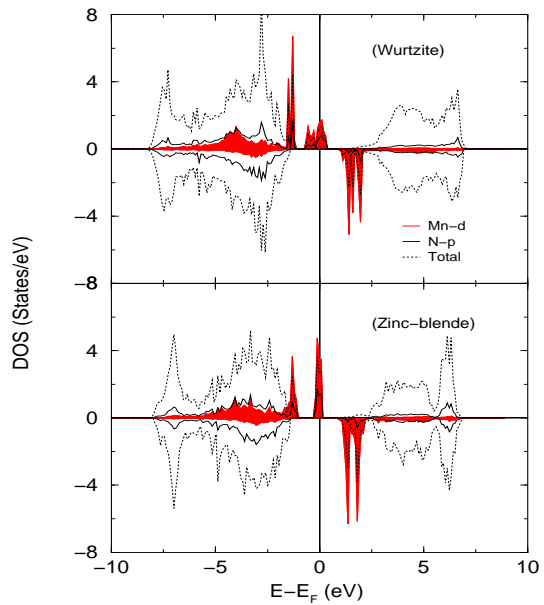}
\caption{Spin resolved density of states of (Mn,Ga)N for (top) wurtzite and (b) zinc-blende
structures. Here, the Mn concentration is 6.25 $\%$.} 
\end{figure}
In Table I, we present the local charges and magnetic moments of Mn and N atoms for different concentrations
of Mn. Charges and magnetic moments of Mn remain almost the same with the concentration variation of Mn. 
The insensitivity of magnetic moment with concentration
 is a signature of localized d-states of Mn. 
 The total moment/cell is always 4.0 $\mu_{B}$ which is the signature of a half-metallic solution. 
The only exception is the case of $x$=0.5, where the total moment/cell is
2.77 $\mu_{B}$. The exchange splitting of Mn d-states is less in this case allowing both spin-up
and spin-down d-states to cross the Fermi level. 
  The averaged induced moments on nearest neighbor N
atoms are also tabulated. In most of the cases, the moments are parallel to Mn moments which is not
the case for a Mn doped GaAs system where nearest neighbor As moments are antiferromagnetically
coupled to the Mn magnetic moment. 

As GaN can be grown in both zinc-blende and wurtzite structures, we have also done calculations
for a 6.5 $\%$ Mn doped GaN system in zinc-blende structure. In Fig. 2, DOSs for both structures
are shown. DOS for a zinc-blende structure is similar to that calculated by Kulatov {\it et al.}
\cite{kulatov}. Also the magnetic moment on Mn atom (3.4 $\mu_{B}$) agrees very well. 
There is no striking difference in the broad features of the DOSs for the two structures. In the 
zinc-blende structure, the peak at the Fermi level is sharper and the nature of these states are
different ($t_{2}$ compared to $e$ states for the wurtzite structure). See fig. 5 and related discussions. 
For both the structures, the Mn impurity peak in the energy gap of the host is separated from
the valence band of the host.
\subsection{LSDA+U calculations}

\begin{figure}[!hbp]
\includegraphics[width=0.45\textwidth,height=0.35\textheight,angle=270]{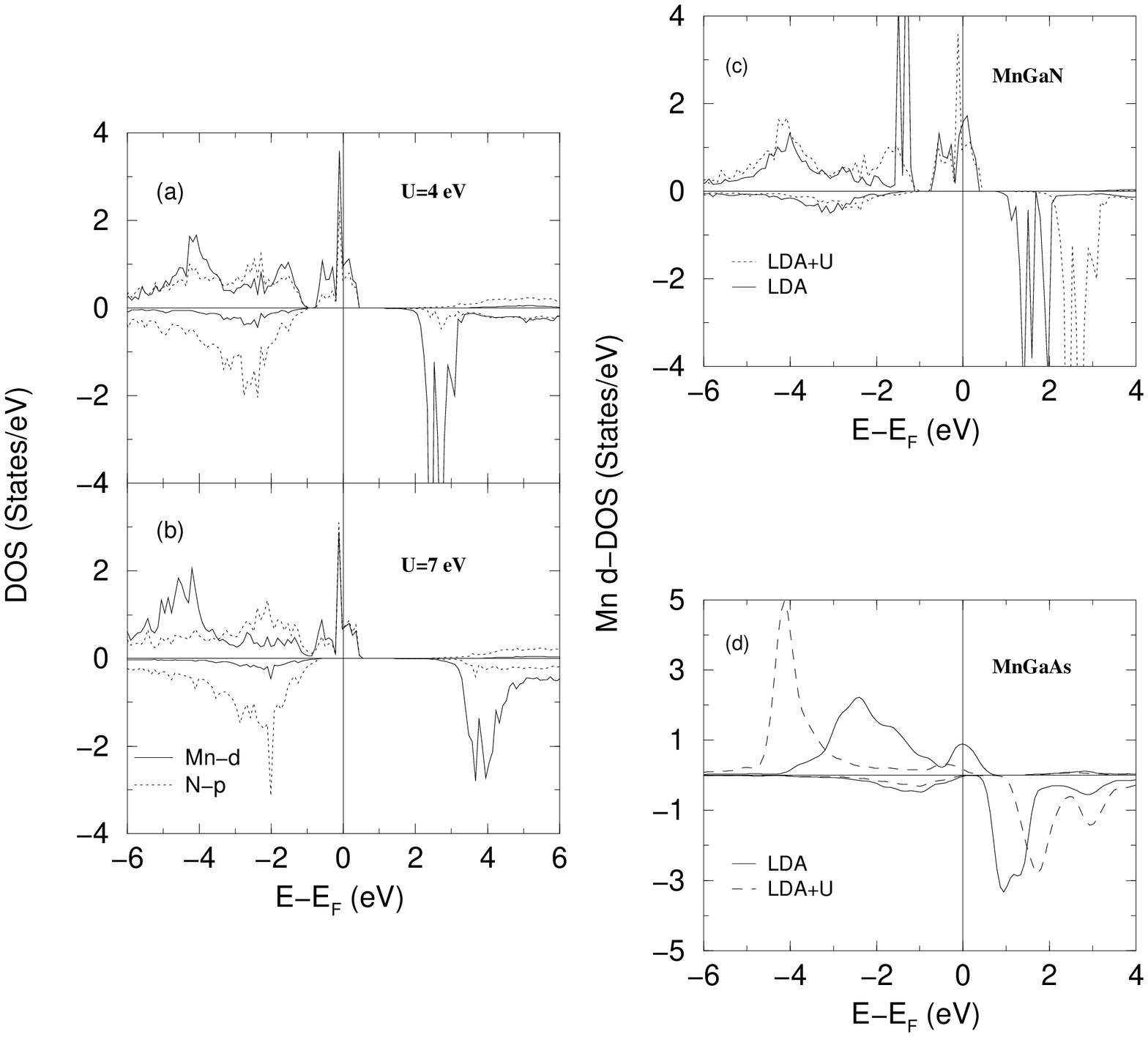}
\caption{(a) and (b), Spin resolved Mn $d$ and N $p$ DOSs from the LSDA+U calculations for different 
values of U for (Mn,Ga)N system in the wurtzite structure, 
 (c) Mn $d$-DOS from both LSDA and LSDA+U calculations for (Mn,Ga)N in the wurtzite structure,
(d) same as (c), but for (Mn,Ga)As in the zinc-blende structure. For all plots, Mn concentration is 6.25 $\%$.}
\end{figure}

It is a matter of debate whether the itinerant band model or the localized atomic model
is appropriate for the description of Mn doped semiconductors. 
Density functional calculations \cite{sanvito,freeman,msm}
based on LSDA or GGA provide the basis of itinerant picture whereas others models are based on a  
localized atomic picture \cite{dietl}. 
 A recent photoemission experiment \cite{kanski} 
on Mn doped GaAs system revealed
the main Mn d-peak to be situated 3.4 eV below the Fermi level.  
 A previous photoemission experiment \cite{okabayashi} reported the 
 peak to be 4.4 eV below the Fermi level. Also, the importance of electron correlation 
 effects in these systems was highlighted. 
Park {\it et. al.} \cite{park} argued from LSDA+U calculations that
correlation corrections are important to have a better agreement with photoemission spectra. 
On the other hand, all density functional calculations based
on LSDA show a peak around 2.6-2.9 eV below the Fermi level \cite{msm,sanvito}.
So it can be argued that the completely localized picture or the completely
itinerant picture cannot solely describe these systems satisfactorily.
\begin{center}
\begin{table}
\caption{Projected charges and
magnetic moments (in $\mu_{B}$) calculated within LSDA+U. The notations are same as in
Table I.}
\begin{tabular}{|l|l|l|l|l|}
\hline \hline 
                                 & U=4.0 eV & U=5.0 eV & U=6.0 eV & U=7.0 eV \\
\hline
Mn-d$^{Q}$			 &  5.01    & 5.00     &  4.99    & 4.98     \\
\hline
Mn$_{mom}$                       &  3.72    & 3.83     &  3.93    & 4.02     \\
\hline
N$_{mom}^{nn}$			 &  -0.02   & -0.04    & -0.05    & -0.06    \\
\hline \hline
\end{tabular}
\end{table}
\end{center}
We have done LSDA+U calculations to investigate the effect of electron correlations.
Firstly, these calculations do not exist in literature and secondly, it is interesting to compare
this simple technique with more rigorous SIC calculations.
In Fig. 3(a-d), we show the DOSs obtained from LSDA and LSDA+U calculations for 6.25 $\%$ Mn.
 As the value of U for Mn $d$ states
is not obtained self-consistently from first principles calculations, we varied U from 4 eV to 7 eV
treating it as a parameter. 
In all cases, the exchange parameter J was considered to be 1.0 eV. Increasing U results in a 
 slow shift of the spin-up impurity band towards the valence band. The small peak around 
1.5 eV below the Fermi level ($E_{F}$) is diminished gradually in magnitude whereas the peak around -4.5 eV
below $E_{F}$ increases in magnitude. Even for U=10 eV (not shown here), the localized peak remains pinned
 close to $E_{F}$ and is {\it not} merged with the delocalized valence band states. 
 The spin down
DOS shifts almost rigidly away from the Fermi level towards higher energy with increasing U. 
Local charges and magnetic moments of Mn and nearest neighbor N atoms are 
listed in Table II. Magnetic moment of Mn increases with U due to increase in localization of $d$ states. 
The induced moment of N atoms also increase. In Fig. 3(c), a comparison between 
 LSDA and LSDA+U calculations is shown. The redistribution of weights of Mn $d$ peaks
with the inclusion of U is visible. In LSDA+U result, the sharp peak is very close to the Fermi level in
the impurity band whereas it is in the valence band in the LSDA calculation. This peak is of 
$e$-character and doesn't take part in the bonding with the neighboring N atoms. These results are in 
agreement with the more sophisticated 
SIC calculations reported by  Filippetti {\it et al.} \cite{alessio}. They also found 
a flat band with $d_{z^{2}}$ character at $E_{F}$. The nature of the states close to $E_{F}$ from 
our calculation is shown later in fig. 5. To our knowledge, the valence band photoemission spectra of
(Mn,Ga)N is not available in the literature. So, the extent of validity of LSDA approach cannot be tested. 
Our LSDA+U findings can be compared with future angle-resolved photoemission experiments on (Mn,Ga)N to 
verify the existence of the localized peak close to $E_{F}$.
A comparative
study with Mn doped GaAs is shown in Fig. 3(d). The main broad Mn $d$ peak around 2.8 eV below the Fermi level
in an LSDA calculation is shifted 4 eV below the Fermi level with a smaller band width. The DOS at the Fermi
level is also decreased compared to that of an LSDA calculation. In both LSDA and LSDA+U calculations, 
the hybridization between Mn $d$ and As $p$ states are seen. So the holes in the valence band have hybridized
$p-d$ character. 
\begin{figure}[!htp]
\includegraphics[width=0.3\textwidth,height=0.38\textheight,angle=270]{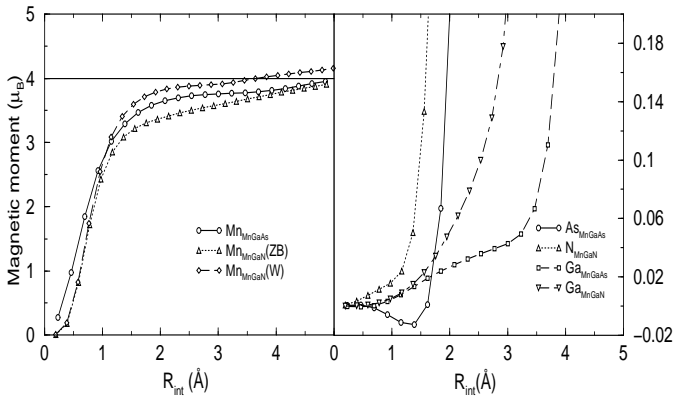}
\caption{Magnetic moment vs. radius of integration around specific atoms. In (a) integrated
magnetization density around Mn spheres and in (b) same around nearest neighbor anions and 
next nearest neighbor Ga atoms are shown. See text for details. The horizontal line in the 
left figure indicates a magnetic moment of 4 $\mu_{B}$. ZB and W represent zinc-blende and
wurtzite structures.}
\end{figure}

In Fig. 4, we show the integrated magnetization density around various atoms in the unit cell
as a function of the radius of integration. It has been calculated as 
$$ M(R) = \int_{0}^{R} (\rho^{\uparrow}(r)-\rho^{\downarrow}(r))dr $$ 
where $\rho^{\uparrow}(r)$ and $\rho^{\downarrow}(r)$ are the spin-up and spin-down charge densities
 respectively and $M$ is the magnetic moment obtained for a radius $R$.
For (Mn,Ga)N and (Mn,Ga)As in the zinc-blende structure, the magnetic moment reaches the value 4 $\mu_{B}$
at a distance of 5 \AA\ far from the Mn center. But, in the wurtzite structure of (Mn,Ga)N, this value is
reached at a smaller distance. This again shows a more localized character of Mn $d$ states in wurtzite 
GaN. The integration around nearest neighbor anions reveal that N has a positive
contribution in (Mn,Ga)N whereas, As has a negative contribution.

\begin{figure}[!hbp]
\includegraphics[width=0.35\textwidth,height=0.38\textheight,angle=270]{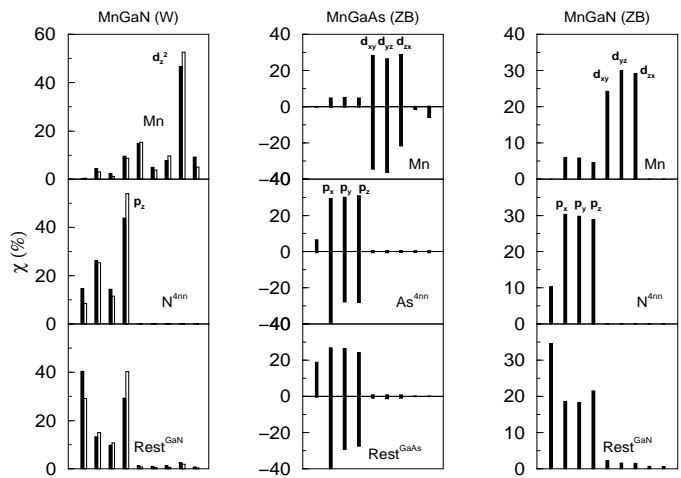}
\caption{Character of states within an energy interval $E_{F}$ to 
$E_{F}$-0.5 eV for (left) (Mn,Ga)N in the wurtzite structure, (middle) (Mn,Ga)As in the
zinc-blende structure and (right) (Mn,Ga)N in the zinc-blende structure having 6.25 $\%$ Mn
in all cases. 
In x-axis, the plotted states are in the order from
left : $s, p_{x}, p_{y}, p_{z}, d_{xy}, d_{yz}, d_{zx}, d_{z^{2}}, d_{x^{2}-y^{2}}$. 
 See text for details. For the middle panel, both spin up and spin down contributions are shown
whereas for the left and right panels, there is no state in the spin down channel. In the left column, 
the empty bars represent LSDA+U values. For each panel, data are shown for Mn(top), four nearest neighbor (nn) 
anions (middle) and the collective contributions from all other
atoms (bottom) in the unit cell.}
\end{figure}

\begin{figure}[!htp]
\includegraphics[width=0.35\textwidth,height=0.35\textheight,angle=270]{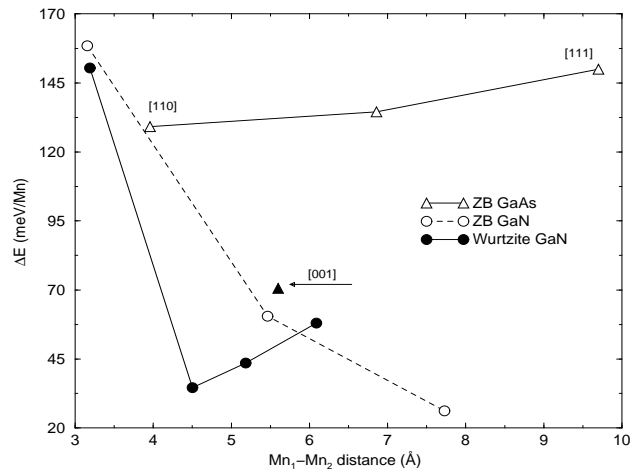}
\caption{Total energy difference $\Delta E$ 
between ferromagnetic and antiferromagnetic configurations vs. Mn-Mn distance in the unit cell.
Data for zinc-blende(ZB) GaN, ZB GaAs and wurtzite (W) GaN are shown. The filled triangle represents
$\Delta E$ for [001] direction in ZB GaAs.}
\end{figure}
In Fig. 5, we show the character of the states within an energy interval close to the Fermi level.
The states close to $E_{F}$ in the electronic structure are important in characterizing the origin
of ferromagnetism. 
The projection of the wavefunction onto spherical harmonics around each atom has been calculated
as mentioned in ref. \cite{eichler}. In the figure, $\chi$ is defined as :
$\chi^{N,\sigma}_{lm} = \sum_{nk} \vert \langle Y^{N,\sigma}_{lm} \vert \phi^{\sigma}_{nk} \rangle \vert^{2}$,
where $\sigma$ and $N$ are the spin and atom indices respectively and $n$ is the index for 
the bands within the specified energy interval. It is clear from the figure that for (Mn,Ga)N
in the wurtzite structure (left panel),
the $d_{z^{2}}$ component of Mn $e$-orbital and the $p_{z}$ component of nearest neighbor 
N p orbital are the dominant states
close to $E_{F}$. On the other hand, Mn $t_{2}$ and nearest neighbor As $p$ states hybridize strongly in (Mn,Ga)As.  
This is also true for (Mn,Ga)N in the zinc-blende structure. Symmetry of the crystal structure and the splitting of
d-states under the corresponding crystal fields determine the position of $t_{2}$ and $e$ states. For the rest of the
atoms in the unit cell, states having $s$ and $p$ character are the dominant.
\section{Interatomic exchange interactions}
To determine the interatomic exchange interactions, we followed a simple model. In the unit cell, two Mn atoms
were placed in various positions. For each Mn-Mn separation,  
ferromagnetic (FM) and antiferromagnetic (AFM) alignments of Mn spins were
considered.  The total energy difference $\Delta E$ ($\Delta E$=$E_{tot}^{AFM}$-$E_{tot}^{FM}$) 
between these two alignments is a measure of interatomic exchange interaction.
In Fig. 6, we plot $\Delta E$ as a function of Mn-Mn separation $d$ for both Mn doped GaAs and Mn doped GaN
systems. For Mn doped GaN, both zinc-blende and wurtzite structures were
considered for these calculations with two Mn atoms in unit cells having 64 atoms and 72 atoms
respectively. For Mn doped GaAs, a 64 atom unit cell was chosen. 
Ferromagnetic interaction between Mn spins is favored for all the cases considered here.
In (Mn,Ga)N, the first nearest neighbor (nn) exchange interaction is the strongest.
The value of $\Delta E$ is increased a little bit compared to the same for (Mn,Ga)As. So, for a defect-free
calculation, there is no indication that (Mn,Ga)N should have a much higher $T_{C}$ than (Mn,Ga)As. 
It suggests that the formation of other phases during the growth is responsible for very high $T_{C}$ observed
in certain experiments. 
It is also seen from the figure that $\Delta E$ decreases sharply with $d$ for 
Mn doped GaN systems. It shows that ferromagnetic exchange interaction in (Mn,Ga)N is short-ranged. 
 This indicates that the formation of Mn clusters within a short radial distance might 
 lead to high values of $T_{C}$.  
For Mn doped GaAs, the exchange interaction is long ranged and doesn't decrease
rapidly. The proper range of ferromagnetic interactions can be studied with a bigger supercell. 
In summary, the results for exchange interactions indicate that the ferromagnetic interaction between
Mn spins in (Mn,Ga)As is mediated by delocalized valence band holes whereas the origin of ferromagnetism
in (Mn,Ga)N may result from a double-exchange mechanism \cite{sato} 
involving the hopping of Mn $d$ electrons. The other probable mechanism can be the formation of
Zhang-Rice magnetic polaron \cite{zr}. 

Another interesting observation is the anisotropy of the exchange interactions in different
crystallographic directions. In (Mn,Ga)As, ferromagnetic coupling is stronger either in the bonding direction e.g. [111]
or in a direction where the two Mn spins are connected by As bonds e.g. in [110] direction. The
coupling decreases for the [001] direction where there is no As atom in between (shown as a filled 
triangle in fig. 6).
 The strong ferromagnetic interactions in [110] or [111]
directions result from strong $p-d$ hybridization between Mn $d$ and As $p$ states. As the interaction is mediated by
the delocalized states, they are sufficiently long-ranged. In [001] direction, the exchange 
interaction between the Mn spins can not be mediated by the $p-d$ hybridization. So, the value of $\Delta E$ 
is decreased compared to the other cases. In (Mn,Ga)N, the exchange interaction
is not mediated by the delocalized valence band states and the ferromagnetic interactions decrease sharply 
with the Mn-Mn separation. So, the anisotropy in the exchange interactions is not significantly observed. We
can conclude that in general, this anisotropy should be present for all Mn doped semiconductors where ferromagnetic
long range interaction is mediated by delocalized valence band states. 
We also studied the exchange interactions in (Mn,Ga)N within the LSDA+U scheme 
in a similar way described above. U=4 eV and U=7 eV
were considered for the calculations. We found that $\Delta E$ for nearest neighbor Mn-Mn distance 
increases from 170 meV/Mn to 240 meV/Mn while going 
from U=4 eV to U=7 eV. Ferromagnetic interactions between Mn spins become stronger as the localization 
of the Mn $d$-states is increased.
\section{Conclusion}
We have studied the electronic structure and magnetism of Mn doped GaN systems for a wide
concentration range of Mn. The deep acceptor level of Mn lies distinctly in the gap of GaN, separated
from the valence band of host GaN. This is in contrast to the case of (Mn,Ga)As where the Mn forms
shallow acceptor level close to the valence band of GaAs.
 Ferromagnetic interactions are short-ranged in (Mn,Ga)N systems whereas they have delocalized 
itinerant character for (Mn,Ga)As. Also, in (Mn,Ga)As, exchange
interactions are significantly anisotropic in different crystallographic directions. This has been explained in terms
of anisotropic $p-d$ hybridization. The results presented here are for ideal systems having no contribution from
the native defects formed during the non-equilibrium growth. To have a more realistic picture, one should take into
account these effects. Also, Mn can occupy interstitial positions in the lattice and the spin interactions
can become quite complicated. We are presently calculating the formation energies of the defects and subsequently the
magnetic interactions in presence of them. Results will be reported in future communications. 

\acknowledgements
We are grateful to the Swedish Foundation for Strategic Research (SSF)
, the Swedish Natural Science Research Council (NFR), and the G\"{o}ran
Gustafsson Foundation for financial
support. We thank J. Hafner and G. Kresse for giving us the possibility to use
VASP. B.S. is grateful to Dr. Lars Nordstr\"om and Prof. Olle Eriksson for valuable suggestions.
 Also, we acknowledge computational support from National Supercomputer Center (NSC) at Link\"oping, Sweden.

\end{document}